\documentclass{aa}
\usepackage{txfonts}
\usepackage{graphicx}
\usepackage{natbib}
\bibpunct{(}{)}{;}{a}{}{,}
\begin{document}
\title{Orbits in the T Tauri triple system observed with SPHERE%
  \thanks{Based on observations collected at the European Southern
    Observatory, Chile, proposals number 070.C-0162, 072.C-0593,
    074.C-0699, 074.C-0396, 078.C-0386, 380.C-0179, 382.C-0324,
    60.A-9363 and 60.A-9364}}
\author{R.\ K\"ohler\inst{1,2}
	\and
	M.\ Kasper\inst{3}
	\and
	T.\ M.\ Herbst\inst{4}
	\and
	T. Ratzka\inst{5}
	\and
        G.\ H.-M.\ Bertrang\inst{6,7,8,9}
}
\offprints{Rainer K\"ohler, \email{rainer.koehler@univie.ac.at}}
\institute{%
	University of Innsbruck, Institute for Astro- and Particle
        Physics, Technikerstr.~25, 6020 Innsbruck, Austria
\and
	University of Vienna, Department of Astrophysics,
        T\"urkenschanzstr.~17 (Sternwarte), 1180 Vienna, Austria
\and
	European Southern Observatory,
	Karl-Schwarzschild-Str.\ 2, 85748 Garching bei M\"unchen, Germany
\and
	Max-Planck-Institut f\"ur Astronomie,
	K\"onigstuhl 17, 69117 Heidelberg, Germany
\and
	University of Graz, Institute for Physics/IGAM, NAWI Graz,
        Universit\"atsplatz 5/II, 8010 Graz, Austria
\and
	Kiel University, Institute of Theoretical Physics and
        Astrophysics, Leibnizstr.~15, 24098 Kiel, Germany
\and
	Departamento de Astronom{\'\i}a, Universidad de Chile, Casilla
        36-D, Santiago, Chile
\and
	Facultad de Ingenier{\'\i}a, Universidad Diego Portales,
        Av.~Ej\'ercito 441, Santiago, Chile
\and
	Millennium Nucleus ``Protoplanetary Disks in ALMA Early
        Science'', Universidad de Chile, Casilla 36-D, Santiago, Chile
}
\date{Received August 5, 2015 / Accepted December 7, 2015}
%%
%%%%%%%%%%%%%%%%%%%%%%%%%%%%%%%%%%%%%%%%%%%%%%%%%%%%%%%%%%%%%%%%%%%%%%%%%%%%%
%%
\abstract{}
{We present new astrometric measurements of the components in the
  T~Tauri system, and derive new orbits and masses.}
{T~Tauri was observed during the science verification time of the new
  extreme adaptive optics facility SPHERE at the VLT.
  We combine the new positions with recalibrated NACO-measurements and
  data from the literature.  Model fits for the orbits of T~Tau~Sa and
  Sb around each other and around T~Tau~N yield orbital elements and
  individual masses of the stars Sa and Sb.}
{Our new orbit for T~Tau~Sa/Sb is in good agreement with other recent
  results, which indicates that enough of the orbit has been observed
  for a reliable fit.  The total mass of T~Tau~S is
  $2.65\pm0.11\,M_\odot$.  The mass ratio $M_{\rm Sb}:M_{\rm Sa}$ is
  $0.25\pm0.03$, which yields individual masses of $M_{\rm Sa} =
  2.12\pm0.10\,M_\odot$ and $M_{\rm Sb} = 0.53\pm0.06\,M_\odot$.
  If our current knowledge of the orbital motions is used to compute
  the position of the southern radio source in the T~Tauri system,
  then we find no evidence for the proposed dramatic change in its
  path.}
{}
\maketitle
%
%%%%%%%%%%%%%%%%%%%%%%%%%%%%%%%%%%%%%%%%%%%%%%%%%%%%%%%%%%%%%%%%%%%%%%%%%%%%%

\section{Introduction}

T~Tauri is a triple star which became the eponymous member of the
class of low-mass, pre-main-sequence stars.
It is located in the Taurus-Auriga star-forming region at a distance
of $146.7\pm0.6\rm\,pc$ \citep{loinard07b}.
As brightest member of the class, it was the subject of many studies
of the stellar components and/or the circumstellar material surrounding
them.

Using speckle-interferometry, \citet{dyck82} were able to resolve
T~Tauri into two components separated by about $0\farcs7$, which are
commonly called T~Tau~N and T~Tau~S.
The southern component is only visible in the infrared, it has not
been detected in the V-band down to 19.6\,mag \citep{stapelfeldt98}.
With the construction of larger telescopes, it became possible to
resolve T~Tau~S itself into two stars named T~Tau~Sa and T~Tau~Sb
\citep{koresko00}.
Because of their small separation ($\sim\,$50\,mas at the time of
the discovery) and short orbital period, it is possible to determine
their orbital parameters within a reasonable time span.
A number of authors have presented orbit fits
\citep[e.g.][]{beck04,johnston04b,duchene06,schaefer06,koehler08a,koehler08b,schaefer14},
where the precision of the orbital parameters improved as more
observational data became available.
%% precision is repeatability of measurements,
%% accuracy is proximity to the true value
The nearby third component T~Tau~N can be used as astrometric
reference to determine the individual orbits of Sa and Sb around their
center of mass \citep{duchene06,koehler08a,koehler08b}.
The ratio of the semi-major axes is equal to their mass ratio, while
the sum of both masses can be computed from the Sa/Sb orbit.
Together, we obtain individual masses for Sa and Sb.
It turned out that T~Tau~Sa is the most massive component in the
triple system, despite its faintness at optical and near-IR
wavelengths (it was never detected in J-band, including the new SPHERE
observations).
The currently most likely explanation for this is that T~Tau~Sa and Sb
are hidden behind circumstellar and/or circumbinary material
\citep{duchene05}.

In this work, we report on our astrometric analysis of observations
taken during the science verification time of SPHERE.
The data was analyzed independently by a different team, who presented
their results in \citet{csepany15}.
In addition to the SPHERE data, we also recalibrated all the
available NACO-data, in order to provide a more consistent astrometric
reference frame.
This includes two NACO-observations that were not published before.
We use these data to derive improved orbital elements and masses,
and discuss the implications for our understanding of the system.

This paper reports on the astrometric measurements taken with SPHERE.
The science verification data also allowed us to image the extended
emission in the T~Tauri system in unprecedented detail. These results
will be presented in a subsequent paper (Kasper et al., submitted to
A\&A).

%%%%%%%%%%%%%%%%%%%%%%%%%%%%%%%%%%%%%%%%%%%%%%%%%%%%%%%%%%%%%%%%%%%%%%%%%%%%%

\begin{table}[t]
 \caption{Pixel Scale and Orientation of our NACO-observations.}
 \label{Table-Cal}
 \begin{center}
  \begin{tabular}{lcc}
    \noalign{\vskip1pt\hrule\vskip1pt}
    Epoch	& Pixel scale		& P.A.\ of the $y$-axis\\
		& [mas/pixel]		& [$^\circ$]\\
    \noalign{\vskip1pt\hrule\vskip1pt}
    Dec.~2001	& $13.241\pm0.011$	& $1.26\pm0.10$	\\
    Dec.~2002	& $13.239\pm0.028$	& $0.01\pm0.06$	\\
    Dec.~2003	& $13.262\pm0.025$	& $0.18\pm0.13$	\\
    Dec.~2004	& $13.295\pm0.024$	& $0.01\pm0.10$	\\
    Dec.~2006	& $13.258\pm0.020$	& $0.50\pm0.06$	\\
    Sep.~2007	& $13.270\pm0.010$	& $0.43\pm0.12$	\\
    Feb.~2008	& $13.283\pm0.004$	& $0.47\pm0.10$	\\
    Oct.~2008	& $13.269\pm0.015$	& $0.54\pm0.10$	\\
    Feb.~2009	& $13.266\pm0.013$	& $0.54\pm0.08$	\\
    \noalign{\vskip1pt\hrule\vskip1pt}
  \end{tabular}
 \end{center}
\end{table}

%\input Table-Obs.tex
%% Table-Obs.tex
%% created by Table-Obs.pro on Thu Mar 12 16:51:45 2015
%% added Filter by hand - RK 27jul2015
%%
\begin{table*}
\def\1{\phantom{1}}

\def\fnote#1{^{\mathrm{#1}}}
\caption{New and recalibrated astrometric measurements of T~Tau N -- Sa -- Sb}
\label{Table-Obs}
%       date,Epoch|sep       esepPA         ePa|sep       esepPA         ePA|src
\begin{tabular}{cccr@{${}\pm{}$}lr@{${}\pm{}$}lcr@{${}\pm{}$}lr@{${}\pm{}$}lcc}
\noalign{\vskip1pt\hrule\vskip1pt}
Date (UT) & Epoch && \multicolumn{4}{c}{T Tau N - Sa} && \multicolumn{4}{c}{T Tau Sa - Sb} & Filter & Instrument \\
	  &	  && \multicolumn{2}{c}{$d$ [mas]} & \multicolumn{2}{c}{P.A.~$[^\circ]$}
						&& \multicolumn{2}{c}{$d$ [mas]} & \multicolumn{2}{c}{P.A.~$[^\circ]$} \\
\noalign{\vskip1pt\hrule\vskip1pt}
2001 Dec 08,  3:34 &   2001.93470  && $693.70$&$ 0.66$	& $181.88$&$0.11$	&& $ 99.60$&$ 0.80$  & $276.80$&$0.40$	& K  & NACO\\
2002 Dec 15,  2:31 &   2002.95306  && $694.21$&$ 1.57$	& $182.56$&$0.10$	&& $106.50$&$ 1.60$  & $284.10$&$0.20$	& Ks & NACO\\
2003 Dec 12,  3:14 &   2003.94424  && $692.93$&$ 2.18$	& $183.07$&$0.13$	&& $112.70$&$ 0.80$  & $288.70$&$0.80$	& Ks & NACO\\
2004 Dec 09,  4:06 &   2004.93818  && $697.89$&$ 1.64$	& $183.74$&$0.14$	&& $119.70$&$ 1.20$  & $296.10$&$0.60$	& Ks & NACO\\
2006 Oct 11,  8:50 &   2006.77582  && $694.75$&$ 1.38$	& $185.21$&$0.07$	&& $125.10$&$ 0.80$  & $304.99$&$0.20$	& Ks & NACO\\
2007 Sep 16,  8:00 &   2007.70659  && $694.41$&$ 0.92$	& $185.59$&$0.13$	&& $127.20$&$ 0.60$  & $308.80$&$0.40$	& Ks & NACO\\
2008 Feb 01,  0:38 &   2008.08358  && $695.10$&$ 0.24$	& $185.95$&$0.10$	&& $127.10$&$ 0.20$  & $310.85$&$0.14$	& Ks & NACO\\
2008 Oct 17,  6:18 &   2008.79333  && $695.00$&$ 1.66$	& $186.36$&$0.10$	&& $129.30$&$ 1.30$  & $314.50$&$0.70$	& Ks & NACO\\
2009 Feb 18,  0:32 &   2009.13216  && $690.84$&$ 2.32$	& $186.94$&$0.13$	&& $124.60$&$ 1.20$  & $315.11$&$0.60$	& Ks & NACO\\
2014 Dec 09,  3:36 &   2014.93676  && $689.98$&$ 0.64$	& $191.53$&$0.10$	&& $110.33$&$ 0.25$  & $345.40$&$0.15$	& BrG & SPHERE\\
2015 Jan 23,  2:20 &   2015.05981  && \multicolumn{2}{c}{---} & \multicolumn{2}{c}{---}	&& $109.27$&$ 0.44$ & $345.70$&$0.25$ & Ks & SPHERE$\fnote{a}$\\
\noalign{\vskip1pt\hrule\vskip1pt}
\end{tabular}
\begin{list}{}{}
  \item[$^{\mathrm{a}}$] T Tau N hidden behind the coronographic mask.
\end{list}
\end{table*}

%%%%%%%%%%%%%%%%%%%%%%%%%%%%%%%%%%%%%%%%%%%%

\section{Observations and data reduction}

Many measurements of the relative positions of two or three of the
components of T~Tau have been published in the past
\citep{ghez95,roddier00,white01,koresko00,koehler00a,duchene02,furlan03,
  beck04,duchene05,duchene06,schaefer06,koehler08a,koehler08b,schaefer14}.
We use all of these data to solve for the orbit, with the exception of
the points by \citet{mayama06}, because they deviate significantly
from the model and the other measurements.

\subsection{NACO}

Between 2001 and 2009, T~Tauri was observed almost yearly with the
adaptive optics, near-infrared camera NACO at the ESO Very Large
Telescope (VLT) on Cerro Paranal, Chile \citep{Rousset03, lenzen03}.
Most of the data have already been published
\citep{koehler08a,koehler08b}, with the exception of the observations
in October 2008 and February 2009.
However, the astrometric calibration of the published data was derived
from a field near the Trapezium in the Orion Nebula Cluster and
coordinates by \citet{mccaughrean94}.
By now, much better coordinates are available \citep{close12}.
Therefore, we recalibrated all the available NACO-data of T~Tauri.
The results of the new calibration are listed in
Table~\ref{Table-Cal}, the new relative positions of T~Tauri in
Table~\ref{Table-Obs}.
The new calibration resulted on average in a pixel scale that is
0.2\,\% larger and a rotation by 0.3\degr\ compared to the old
calibration.

%%%%%%%%%%%%%%%%%%%%%%%%%%%%%%%%%

\subsection{SPHERE}

T~Tauri was observed during the science verification time of the
extreme adaptive optics facility SPHERE (Spectro-Polarimetric
High-contrast Exoplanet REsearch) at the VLT \citep{beuzit08}
on December 9, 2014 and January 23, 2015.
It was the target of two different science programs
(PIs: Markus Kasper and Gergely Cs\'ep\'any).

\begin{figure}[t]
\centerline{\includegraphics[width=0.9\hsize]{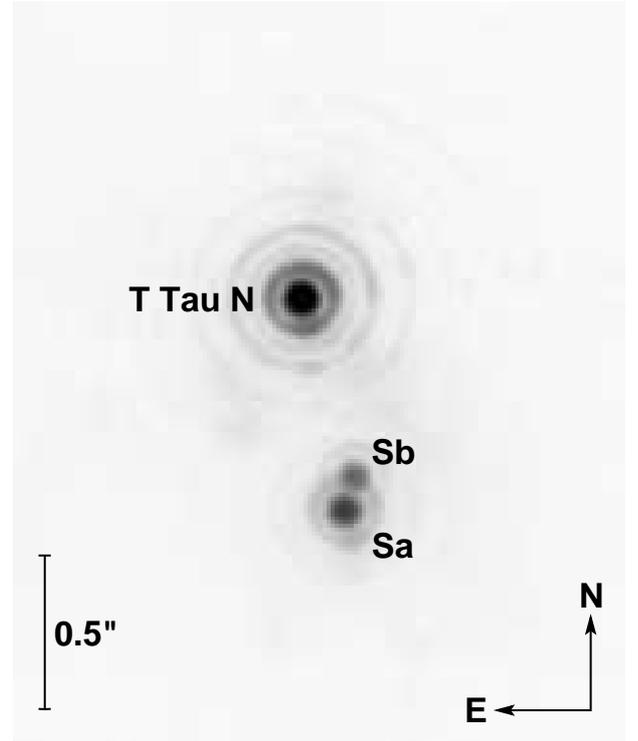}}
\caption{The T~Tauri system imaged with SPHERE on December 9, 2014,
  displayed with a logarithmic scale.}
\label{SpherePic}
\end{figure}

We applied standard methods for infrared data reduction (dark
subtraction, flatfield, and correction of bad pixels).
Figure~\ref{SpherePic} shows one of the reduced images.
The positions of the stars were measured with the Starfinder software
\citep{Diolaiti00}.
Uncertainties were estimated by computing the standard deviation
of positions measured in individual images.

In December 2014, T~Tauri was imaged with the infrared dual-band
imager and spectrograph (IRDIS) through a number of narrow-band
filters.  We measured the relative positions of the three components
of T~Tauri in the images obtained with the BrG filter, which has a
central wavelength of $2162.9\rm\,nm$.
The astrometric calibration was carried out by the instrument team and
is described in the SPHERE user manual.\footnote{Document number
  VLT-MAN-SPH-14690-0430, issue P96.1}
During the science verification campaign, SPHERE observed a dedicated
field in the outer region of the globular cluster 47 Tuc to derive
astrometric calibration of the IRDIS broad-band H image, using the
Hubble Space Telescope (HST) data as a reference (see
\citet{bellini2014} for the methods used to obtain HST
measurements).
The derived plate-scale was $12.232\pm0.005\rm\,mas/pixel$ (for
observations without a coronograph), and the $y$-axis of the IRDIS
detector was at a position angle (P.A.) of $-1.764\pm0.01^\circ$.
The anamorphic distortion was corrected by multiplying the
y-coordinates by $1.0062$.
We used images taken with SPHERE's internal grid mask to confirm that
the pixel scale is the same with the H-band and the BrG filters.

In January 2015, images were recorded with IRDIS through a K-band
filter.
For most of these observations, T~Tau~N was hidden behind a
coronographic mask \citep{csepany15}.
Therefore, we did not measure its position, but only the relative
positions of T~Tau~Sa and Sb.
The pixel scale and orientation provided by the SPHERE consortium were
used for the astrometric calibration.

The results for both observations are presented in
Table~\ref{Table-Obs}.
These data sets were reduced independently by \citet{csepany15}, with
different results for the positions.
At the time of their analysis, the precise pixel scale and true north
orientation were not known, which explains the different positions
(G.\ Cs{\'e}p{\'a}ny, priv.\ comm.).

%%%%%%%%%%%%%%%%%%%%%%%%%%%%%%%%%%%%%%%%%%%%%%%%%%%%%%%%%%%%%%%%%%%%%%%%%%%%%

\section{Orbit determination\label{fit}}

\subsection{The orbit of T Tauri Sa/Sb}
\label{OrbitSaSbSect}

\begin{figure}[t]
\centerline{\includegraphics[width=\hsize]{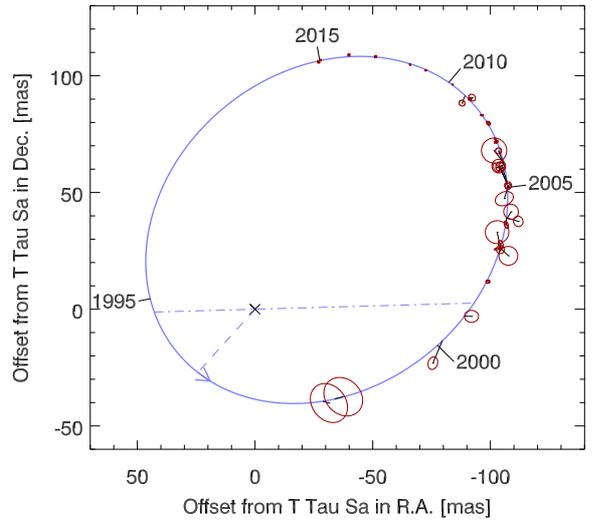}}
\caption{\small Best fitting orbit of T~Tau~Sb around Sa, in the rest frame
  of Sa. The observed positions are marked by their error ellipses and
  lines connecting the observed and calculated position at the
  time of the observations.  The dash-dotted line indicates the line
  of nodes, the dashed line the periastron, and the arrow shows the
  direction of the orbital motion.
}
\label{OrbSabPic}
\end{figure}

To determine the orbit of T~Tau Sa and Sb around each other, we
followed the method described in \citet{koehler08a}.
The first step was to find the best period, eccentricity and time of
periastron with a grid search, while the remaining four orbital
elements were written in the form of the Thiele-Innes constants and
found using singular value decomposition.
In a second step, we improved the results of this grid search by
fitting all seven parameters simultaneously with a
Levenberg-Marquardt\ $\chi^2$ minimization algorithm \citep{press92}.
The orbit with the global minimum $\chi^2$
is shown in Fig.~\ref{OrbSabPic}, and its orbital elements are listed
in Table~\ref{fit2015-3sc-orbitTab}.
%To convert the semi-major axis from mas to AU, we used the
%distance of $146.7\pm0.6\rm\,pc$ \citep{loinard07b}.

%%%%%%%%%%%%%%%%%%%%%%%%%%%%%%%%%%%%%%%%%%%%
%\input{SaSb-fit3-orbit.tex}
%
% Ttau/Paper2015/SaSb-fit3-orbit.tex
% copy of ../Orbit-2015/SaSb/fit2015-3e-orbit.tex
% with some edits
%
% created by lmfinderr on Wed May 27 16:56:38 2015
%
% third attempt with SPHERE, scaled chi^2 for error estimate
%
\begin{table}
\caption{Parameters of the best orbital solution for Sa -- Sb.}
\label{fit2015-3sc-orbitTab}
\label{SaSb-orbitTab}
%% stretch a bit so that sub- and superscripts don't overlap
\renewcommand{\arraystretch}{1.3}
\begin{center}
\begin{tabular}{lr@{}l}
\noalign{\vskip1pt\hrule\vskip1pt}
Orbital Element				& \multicolumn{2}{c}{Value} \\
\noalign{\vskip1pt\hrule\vskip1pt}
Date of periastron $T_0$			& $2450131$ & $\,^{+ 208}_{ -288}$\\
						& (1996 Feb 17)\span\\
Period $P$ (years)                              & $     27$ & $\,^{+   2}_{  -2}$\\
Semi-major axis $a$ (mas)                       & $     85$ & $\,^{+   4}_{  -2}$\\
Semi-major axis $a$ (AU)                        & $   12.5$ & $\,^{+0.6}_{-0.3}$\\
Eccentricity $e$                                & $   0.56$ & $\,^{+0.07}_{-0.09}$\\
Argument of periastron $\omega$ ($^\circ$)      & $     48$ & $\,^{+  34}_{ -25}$\\
P.A. of ascending node $\Omega$ ($^\circ$)      & $     92$ & $\,^{+  26}_{ -36}$\\
Inclination $i$ ($^\circ$)                      & $     20$ & $\,^{+  10}_{  -6}$\\
System mass $M_{\rm S}$ ($\rm mas^3/year^2$)      & $    838$ & $\,^{+  29}_{ -32}$\\
Mass error from distance error ($M_\odot$)	& $	$ & $\pm   0.032$\\
System mass $M_{\rm S}$ ($M_\odot$)                & $   2.65$ & $\,^{+0.10}_{-0.11}$\\
reduced $\chi^2$				& $    2.7$\\
\noalign{\vskip1pt\hrule\vskip1pt}
Mass ratio $M_{\rm Sb}/M_{\rm Sa}$			& $   0.25$ & $\,\pm 0.03$\\
Mass of Sa $M_{\rm Sa}$ ($M_\odot$)		& $   2.12$ & $\,\pm 0.10$\\
Mass of Sb $M_{\rm Sb}$ ($M_\odot$)		& $   0.53$ & $\,\pm 0.06$\\
\noalign{\vskip1pt\hrule\vskip1pt}
\end{tabular}
\end{center}
\end{table}

%%%%%%%%%%%%%%%%%%%%%%%%%%%%%%%%%%%%%%%%%%%%
%%%%%%%%%%%%%%%%%%%%%%%%%%%%%%%%%%%%%%%%%%%%

Uncertainties of the orbital elements were determined by analysing the
$\chi^2$ function around its minimum.
The 68\,\% confidence region for each parameter (the familiar
$1\sigma$-interval in the case of a normal distribution) is the region
where $\chi^2 < \chi^2_{\rm min} + 1$.
Therefore, the uncertainties for the fit parameters depend on the
value of $\chi^2$ at and around the minimum.
The reduced $\chi^2$ of our fit was 2.7, more than one would expect
for a good fit.  This indicated that some of the measurement errors
might have been underestimated.
To avoid underestimating the uncertainties of the orbital elements as
well, we rescaled the uncertainties of the observations so that the
minimum $\chi^2$ was 1.  Although some of the observations showed
larger deviations from the model than others, we had no reason to
trust any of the measurements less than the others.  Therefore, we
multiplied all observational uncertainties by the same factor
$\sqrt{2.7} = 1.64$.
The uncertainties of the orbital elements in
Table~\ref{fit2015-3sc-orbitTab} are based on $\chi^2$ computed
with these scaled measurement uncertainties.

Estimating the uncertainty of the mass required a special procedure.
The mass itself was computed using Kepler's third law ($M=a^3/P^2$,
\citealt{Kepler1619}).
The semi-major axis $a$ and the period $P$ are usually strongly
correlated.  To obtain a realistic estimate for the uncertainty of the
mass, we did {\em not\/} use standard error propagation.
Instead, we considered a set of orbital elements where the
semi-major axis was replaced by the mass.  This is possible because
Kepler's third law gives an unambiguous relation between the two
sets of elements.  With the mass being one of the orbital elements,
we treated it as one of the independent fit parameters and determined
its uncertainty in the same way as for the other parameters.

%%%%%%%%%%%%%%%%%%%%%%%%%%%%%%%%%%%%%%%%%%%%

\subsection{The orbit of T Tauri S and the mass ratio
  $M_{\rm Sb}/M_{\rm Sa}$}
\label{OrbitNSSect}

The orbit of T~Tau Sa and Sb around each other allows us to determine
the sum of the masses of both components.  To measure individual
masses, we need to know the position of the center of mass (CM) of Sa
and Sb.  We followed the method described in \citet{koehler08a} to do
this.  The path of the CM of Sa and Sb can be described in two ways:
It is in orbit around T~Tau N, and it is on the line from Sa to Sb,
where its distance from Sa is a constant fraction of the separation of
Sa and Sb.
The fraction\footnote{The parameter $f$ is often called fractional
  mass \citep{heintz78}, since it is the secondary star's fraction of
  the total mass in a binary.} is $f = q/(1+q)$, with the mass ratio
$q=M_{\rm Sb}/M_{\rm Sa}$.

Our model solves for eight parameters: the seven orbital elements of
the orbit of the CM around T~Tau~N, and the fraction $f$.
We compute $\chi^2$ from the difference between the position on the
orbit and the position of the CM computed from the observed positions
of Sa and Sb (note that the latter depends on the free parameter
$f$).
The fitting procedure is similar to that used for the orbit of Sa/Sb,
except that the grid-search is carried out in 4 dimensions:
eccentricity $e$, period $P$, time of periastron $T_0$, and the
fractional mass~$f$.
Singular value decomposition was used again to fit the Thiele-Innes
constants and hence the remaining orbital elements.
To improve the fit further, about 150 orbit models with $\chi^2$ near
the minimum were selected as starting points for a Levenberg-Marquart
fit.
Since this method assumes that the measurement errors are independent
of the fitting parameters, $f$ cannot be treated like the other
parameters.  Instead, we perform a grid search over a narrow range in
$f$ to find the minimum.

\citet{koehler08a} and \citet{koehler08b} included unresolved
observations of T~Tau~S as approximation for the position of the
center of mass.
However, the unknown offset between the center of light and the
center of mass introduces an additional systematic error.
Therefore, we used only resolved measurements of the positions of
T~Tau~Sa and Sb in this work.
Since both \citet{koehler08b} and this work are based on about 18
years of observational data, we can expect a fit of comparable
quality.

If only the astrometric data were used, the best orbit
solution corresponds to a system mass of 111$\,M_\odot$, which is
clearly unphysical.
In the same way as in \citet{koehler08a}, we introduced a system
  mass of $4.7\pm1.0\,M_\odot$ as an additional constraint for the
  computation of $\chi^2$:
$$
\chi^2 = \sum_i \left(\vec r_{i,\rm model} - \vec r_{i,\rm obs}
			\over \Delta\vec r_{i,\rm obs}\right)^2
	+ \left(M_{\rm model}-4.7\,M_{\odot} \over 1.0\,M_{\odot}\right)^2
$$
Here $\vec r_{i,\rm obs}$ and $\Delta\vec r_{i,\rm obs}$ are the
observed position at time $i$ and the corresponding error, $\vec
r_{i,\rm model}$ is the position at time $i$ computed from the model,
and $M_{\rm model}$ is the system mass computed from the orbit model.
The mass estimate of $4.7\,M_\odot$ is the sum of our result for
$M_{\rm S}$ in Sect.~\ref{OrbitSaSbSect} and the mass of T~Tau~N
estimated from its spectral energy distribution by
\citet{loinard07b}.
They derived a mass for T~Tau~N of $1.83^{+0.20}_{-0.16}$ or
$2.14^{+0.11}_{-0.10}$, depending on the theoretical pre-main-sequence
track used.
The uncertainty of $\pm1.0\,M_\odot$ for our mass estimate was chosen
to be about three times as large as the uncertainties of the estimates
for $M_{\rm S}$ and $M_{\rm N}$,
in order to avoid adding a constraint to the orbit fit that is more
stringent than the astrometric data.

%%%%%%%%%%%%%%%%%%%%%%%%%%%%%%%%%%%%%%%%%%%%
%%\input{NS-fit-orbit.tex}
%
% NS-fit-orbit.tex
% based on fit2015-6fixpe.tex and edited by hand
%
% NOTE: Don't call the system mass M_S !!!
%
% created by TTau/Orbit-2015/N-S/lmfinderrq on Wed Sep  2 23:39:04 2015
%
% all new data, only resolved obs, mass estimate based on Sa/b fit
%
\begin{table}
\caption{Parameters of the best orbital solution for T~Tau N--S.}
\label{NS-fit-orbitTab}
%% stretch a bit so that sub- and superscripts don't overlap
\renewcommand{\arraystretch}{1.3}
\begin{center}
\begin{tabular}{lr@{}l}
\noalign{\vskip1pt\hrule\vskip1pt}
Orbital Element				& \multicolumn{2}{c}{Value} \\
\noalign{\vskip1pt\hrule\vskip1pt}
%Date of periastron $T_0$			& $2439796$ & $\,^{+        9130}_{       -9960}$\\
%						& (1967 Nov  1)\span\\
Date of periastron $T_0$ (Epoch)		& $  1967$ & $\,^{+25}_{-47}$\\
Period $P$ (years)                      	& $   4200$ & $\,^{+5000}_{-3400}$\\
Semi-major axis $a$ (arcsec)            	& $    2.9$ & $\,^{ +5.4}_{ -1.7}$\\
Semi-major axis $a$ (AU)                	& $    430$ & $\,^{+ 790}_{ -250}$\\
Eccentricity $e$                        	& $   0.7$ & $\,^{+ 0.2}_{-0.4}$\\
Argument of periastron $\omega$ ($^\circ$)	& $     12$ & $\,^{+10}_{ -6}$\\
P.A. of ascending node $\Omega$ ($^\circ$)	& $    156$ & $\,^{+11}_{-11}$\\
Inclination $i$ ($^\circ$)                	& $     52$ & $\,^{+ 4}_{ -5}$\\
System mass ($\rm mas^3/year^2$)		& $   1468$ & $\,^{+15}_{-15}$\\
Mass error from distance error ($M_\odot$)	& $	$ & $\pm   0.057$\\
System mass ($M_\odot$)                          & $   4.6$ & $\,^{+ 0.1}_{-0.1}$\\
reduced $\chi^2$				& $    3.7$\\
\noalign{\vskip1pt\hrule\vskip1pt}
\end{tabular}
\end{center}
\end{table}

%%%%%%%%%%%%%%%%%%%%%%%%%%%%%%%%%%%%%%%%%%%%

\begin{figure}[t]
\centerline{\includegraphics[angle=90,width=\hsize]{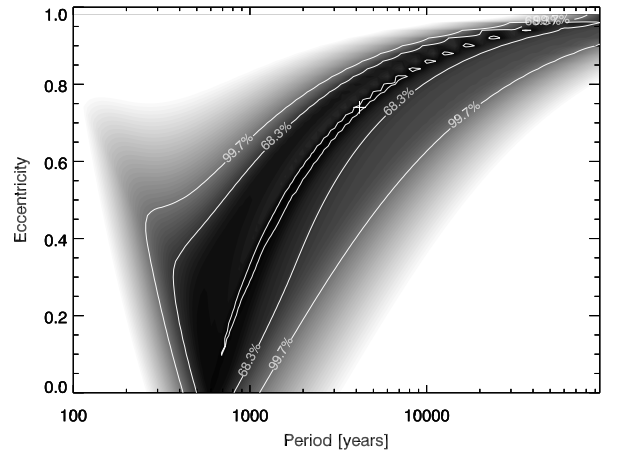}}
\caption{\small $\chi^2$ as function of period and eccentricity for the
  orbit of T~Tau~S around T~Tau~N.
  The cross marks the location of the best-fitting orbit solution (see
  Table~\ref{NS-fit-orbitTab}).
  Contour lines show the $68.3\,\%$ and $99.7\,\%$ confidence
  regions (equivalent to $1\sigma$ and $3\sigma$ for a normal
  distribution).
  The innermost contour line is at $\chi^2_{\rm min} + 0.2$.  This has
  no particular significance, but demonstrates how flat $\chi^2$ is
  around the minimum.}
\label{OrbNSchi}
\end{figure}

%%%%%%%%%%%%%%%%%%%%%%%%%%%%%%%%%%%%%%%%%%%%

\begin{figure}[t]
\centerline{\includegraphics[width=1.1\hsize]{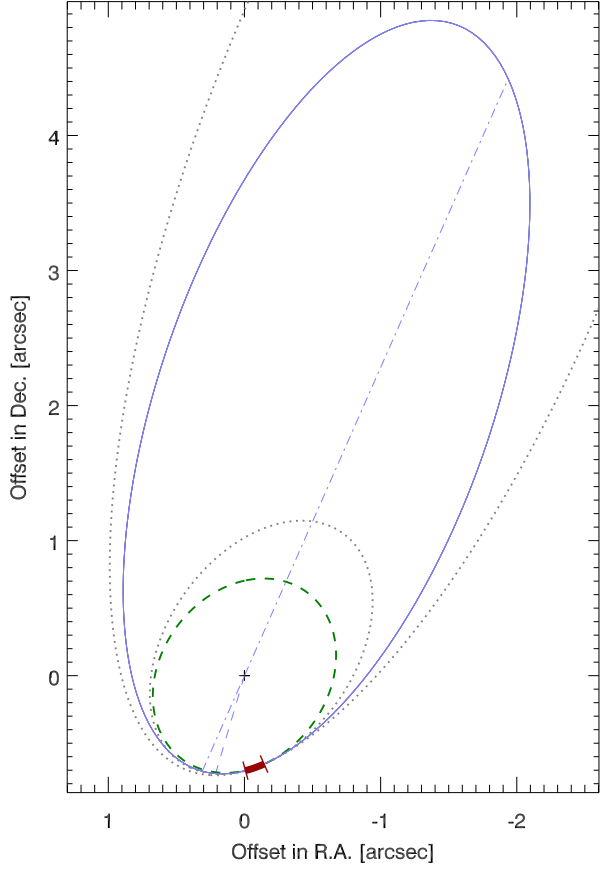}}
\caption{\small Motion of T~Tau~S around T~Tau~N.
  The solid line shows the orbit with the minimum $\chi^2$.
  The dotted lines show orbits at the short and long end of the
  confidence interval for the period (see
  Table~\ref{NS-fit-orbitTab}),
  while the best circular orbit is shown by the dashed line.
  The thick section marks the part of the orbit that has been observed
  so far.}
\label{OrbNSpic}
\end{figure}

The model with minimum $\chi^2$ is presented in
Table~\ref{NS-fit-orbitTab}, while the mass ratio
$M_{\rm Sb}/M_{\rm Sa}$ is given in Table~\ref{fit2015-3sc-orbitTab}.
Uncertainties were determined in the same way as for the orbit of
Sa/Sb, Table~\ref{NS-fit-orbitTab} gives the 68\,\% confidence
intervals.
The uncertainties for the orbital elements are still large, due to the
short section covered by the observations.
Figure~\ref{OrbNSchi} shows $\chi^2$ of our orbit models as function
of period and eccentricity.
The orbit model whose parameters are listed in
Table~\ref{NS-fit-orbitTab} is only one out of a wide range of models
that fit the data almost equally well.
Figure~\ref{OrbNSpic} shows the (formally) best orbit, the orbits at
the lower and upper end of the 68\,\% confidence interval for the
orbital period, and one circular orbit (eccentricity $e=0$).  In the
part of the orbit where observations have been collected, they are
indistinguishable.

\begin{figure}[t]
\centerline{\includegraphics[angle=90,width=\hsize]{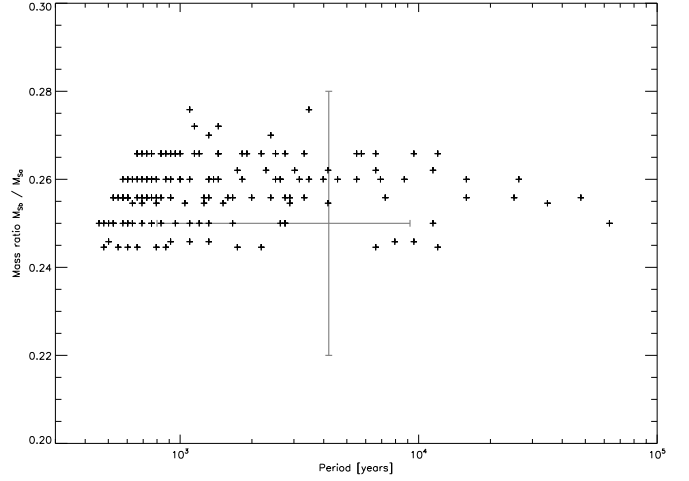}}
\caption{\small Mass ratio $M_{\rm Sb}/M_{\rm Sa}$ as function of the
  period of the T~Tau~N-S orbit model.  These are the results of 150
  runs of a Levenberg-Marquart fit starting with different initial
  estimates for the parameters (see Sect.~\ref{OrbitNSSect}).
  The quantized appearance is caused by the grid-based fit
  method for the mass ratio.
  The large cross indicates the 68\,\% confidence intervals
  for the period and mass ratio (see Tables
  \ref{fit2015-3sc-orbitTab} and \ref{NS-fit-orbitTab}).
  Note that the models plotted encompass a larger range in period
  than the 68\,\% confidence interval of the best solution.
}
\label{MassRatPic}
\end{figure}

However, the main purpose of the fit is to determine the mass ratio of
Sa and Sb, which is well constrained.
Figure~\ref{MassRatPic} shows the mass ratio as function of the
orbital period for 150 orbit models that were the result of
Levenberg-Marquart fits.
Despite the large range in period, the mass ratio of all these
solutions are remarkable similar.
Even though the orbit of T~Tau~S around N is only weakly constrained,
we are confident that our mass ratio $M_{\rm Sb}/M_{\rm Sa}$ is
accurate within the uncertainties.

\begin{figure}[t]
\centerline{\includegraphics[width=\hsize]{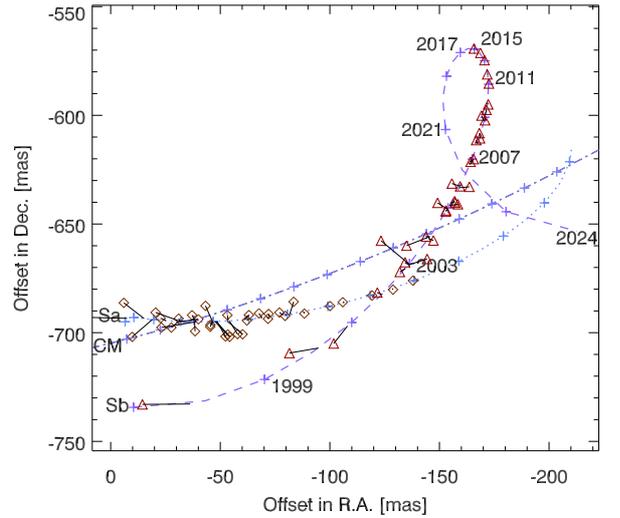}}
\caption{\small Motion of T Tau Sa and Sb in the reference frame of T~Tau~N.
The path of Sa predicted by our orbit models is shown by the dotted
line, the path of Sb by the dashed line, and the path of their center
of mass by the dash-dotted line.
The predicted positions on January 1 in every other year between 1997
and 2024 are marked by crosses.
The observed positions of Sa are depicted by diamonds, those of Sb by
triangles.
Solid lines connect the observed and the predicted positions.}
\label{OrbNSabPic}
\end{figure}

\begin{figure}[t]
\centerline{\includegraphics[width=\hsize]{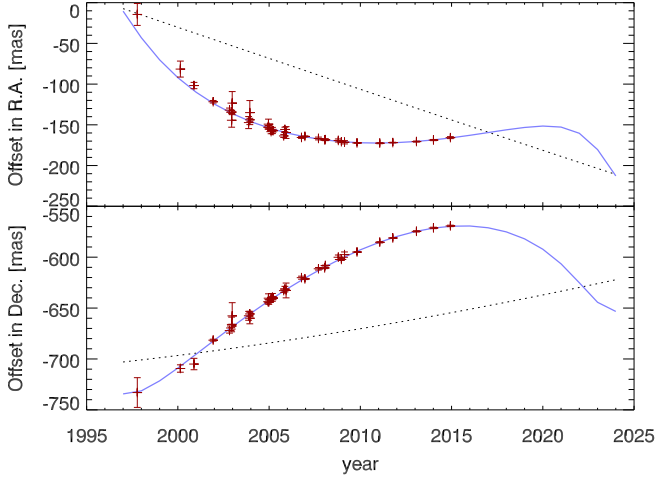}}
\caption{\small Offsets of T~Tau Sb from T~Tau~N.
The observed positions are marked by crosses with error bars.
The solid line is our best fit for the motion of T~Tau~Sb, while the
dotted line shows the motion of the center of mass of T~Tau~Sa and
Sb.}
\label{OrbNSsepPAPic}
\end{figure}

Figure~\ref{OrbNSabPic} shows the motion of T~Tau~Sa and Sb in the
rest frame of T~Tau~N, while Fig.~\ref{OrbNSsepPAPic} shows the
offset of T~Tau~Sb as function of time.  Both figures demonstrate that
our model is a good fit for the motion of Sa and Sb relative to
T~Tau~N.

The high eccentricity of the best-fitting orbit is surprising.
However, high-eccentricity orbits are often the result of orbit fits
to data that cover only a small fraction of the orbit.  As more data
are collected, the orbit solution usually converges towards an orbit
with much lower eccentricity.  We expect that this is also the case
for the orbit of T~Tau~N/S.  Therefore, the orbit presented in
Table~\ref{NS-fit-orbitTab} should not be regarded as the most likely
orbit, but only as one example of a large family of possible
solutions.  Even the confidence intervals in
Table~\ref{NS-fit-orbitTab}, while formally correct, are only an
approximate representation to the probability distribution of the
parameters.  Even a circular orbit can not be ruled out from the data
available (see Fig.~\ref{OrbNSpic}).

%%%%%%%%%%%%%%%%%%%%%%%%%%%%%%%%%%%%%%%%%%%%

%%%%%%%%%%%%%%%%%%%%%%%%%%%%%%%%%%%%%%%%%%%%

  Given the large uncertainty about the orbit, it might be
  surprising that the uncertainty of the system mass is very small,
  significantly smaller than in the term we added in the computation
  of $\chi^2$.
  However, to determine the mass, it is not always necessary to know
  the full orbit \citep[see Sect.~4.2 in][]{koehler08a}.
  The mass can be computed if both separation and acceleration are
  known.  This is usually not the case for astrometric orbits, since
  their component parallel to the line of sight cannot be measured.
  However, these components are zero if the companion is at one of the
  nodes of the orbit.
  The information about separation and acceleration is inherently
  present in the data and results in an orbit solution with a
  well-constrained mass.

  In the case of the orbit of T~Tau~N and S, the mass constraint
  selected one family of orbits from the larger set of orbits that
  agree with the astrometric data.  For this family of orbits, the
  line of nodes is sufficiently close to the observations to result in
  a small mass uncertainty.

%%%%%%%%%%%%%%%%%%%%%%%%%%%%%%%%%%%%%%%%%%%%%%%%%%%%%%%%%%%%%%%%%%%%%%%%%%%%%

\section{Discussion}

\subsection{Comparison to previous results}

The orbital elements for our best fit for the motion of T~Tau~Sa and
Sb around each other are very similar to the preliminary orbit
presented by \citet{schaefer14}.
Both solutions agree well within $1\sigma$.
We take this as sign that the fraction of the orbit observed so far is
large enough to allow a reliable orbit fit, and that our orbits are
close to the true orbit.

  Most of the uncertainties of our orbit are significantly smaller
  than for the orbit of \citet{schaefer14}.  Although more and better
  data should improve the fit, we did not expect to reduce some of the
  uncertainties by a factor of two, since the orbital coverage was
  only extended by one year.  This might be due to the improved
  calibration of the old NACO data.  However, the argument of
  periastron and the P.\,A. of the line of nodes of our solution are
  less well constrained than those of \citet{schaefer14}.
  The position of the periastron is difficult to determine, since the
  inclination of the orbit counteracts the (already moderate)
  eccentricity of the orbit, resulting in a projected orbit that is
  closer to circular than the true orbit.
  The small inclination of only $20^\circ$ (smaller than in
  \citet{schaefer14}) makes it more difficult to find the orientation
  of the line of nodes.  Finally, the P.\,A.\ of the line of nodes and
  the argument of periastron are correlated, since only both angles
  together give the position of the periastron relative to north.

We note that a similar orbit was already published in
\citet{koehler08b}, which was significantly different from the orbit
in \citet{koehler08a}, although only one additional astrometric
measurement was used in \citet{koehler08b}.

For the orbit of T~Tau~N and S around each other, the observed section
is very small (less than 10\degr\ in position angle), which results in
very large errors for any orbital elements derived from the data.
It would be premature to even speak of a ``preliminary orbit''.
Given the large uncertainties, it is also not possible to predict when
a reasonable solution will be possible.
  Nevertheless, our orbit solution is similar to the orbits
  presented in \citet{koehler08b} and \citet{csepany15}.
  The period and semi-major axis are better constrained, due to the
  new data acquired since \citet{koehler08b} and the better
  calibration compared to \citet{csepany15}.

However, to estimate the mass ratio $M_{\rm Sb}/M_{\rm Sa}$, we do not
need the full orbit of T~Tau~S around N, but only a good estimate of
the path of the center of mass at the times of our observations.
Therefore, we can derive the mass ratio with relatively small
uncertainties.
In fact, our value of $0.25\pm0.03$ is close to those reported by
\citet{duchene06}\footnote{Note that their parameter $\mu$ is the
  ratio of $M_{\rm Sb}$ to the total mass $M_{\rm S}$, i.e.\ the same
  as our fractional mass $f$.  Their value of $\mu$ corresponds to
  $q=0.22\pm0.07$.}
and \citet{koehler08b}.
Inclusion of the new data reduced the uncertainty of the mass
ratio and the masses of Sa and Sb by a factor 2 -- 3.

%%%%%%%%%%%%%%%%%%%%%%%%%%%%%%%%%%%%%%%%%%%%%%%%%%%%%%%

\subsection{The compact radio source near T~Tau Sb}

\cite{loinard03,loinard05} observed a compact radio source which they
identified with T~Tau Sb.  Their astrometric measurements indicate a
dramatic change in the motion of the radio source.
\cite{loinard03} speculated that T~Tau Sb underwent an ejection and
might even leave the system on an unbound orbit.
However, \cite{furlan03} found that T~Tau~Sb and the radio source move
on distinct paths, and suggested a fourth object as the source of the
radio emission.

In our orbit fit described in Sect.~\ref{OrbitSaSbSect}, we also
tried parabolic and hyperbolic orbits for T~Tau~Sb, i.e. unbound
orbits that would result in T~Tau~Sb leaving the system.
These orbits do not fit the data very well, the best orbit has a
reduced $\chi^2 = 5.82$, much higher than the best elliptical (bound)
orbit.  Furthermore, if T~Tau Sb was on one of the unbound orbits
matching the data, it would leave the system in an eastern direction.
This is almost the opposite direction of the motion of the radio
source \citep[cf.][]{loinard03}.

\begin{figure}[t]
{\includegraphics[width=0.95\hsize]{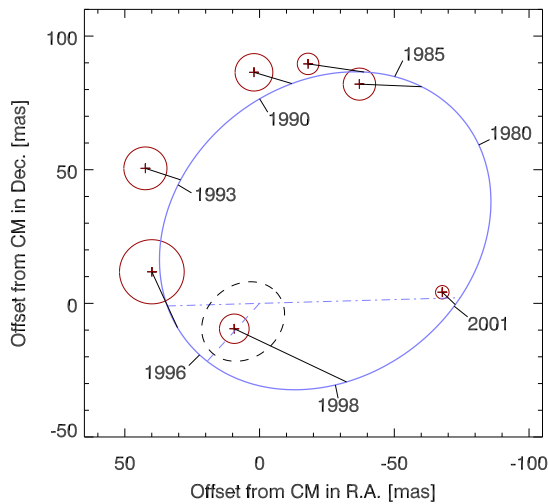}}
\caption{\small Positions of the radio source reported by \citet{loinard03},
  in the reference frame of the center of mass of T~Tau Sa+Sb.
  Overplotted are our orbits for T~Tau~Sa (dashed ellipse) and
  T~Tau~Sb (continuous ellipse).
  Lines connect the positions of the radio source with the expected
  position of T~Tau Sb at the same time.}
\label{LoinardPic}
\end{figure}

In Fig.~\ref{LoinardPic}, we compare the positions of the radio
source reported by \cite{loinard03} and our orbit of T~Tau~Sb, both in
the reference frame of the center of mass of T~Tau~Sa+Sb.
Note that the radio positions in this plot differ from those in
\citet{loinard03}.
\citet{loinard03} computed the offsets from T~Tau~Sa with a linear
approximation of the path of Sa, which does not take into account the
orbital motion of Sa and Sb.
We recalculated the offsets of the radio source from T~Tau~N, and
subtracted the position of the CM of Sa+Sb, computed from the
orbit found in Sect.~\ref{OrbitNSSect}.
With the recalculated offsets, there is hardly any sign of a dramatic
change of its orbit.

%%%%%%%%%%%%%%%%%%%%%%%%%%%%%%%%%%%%%%%%%%%%
%%%%%%%%%%%%%%%%%%%%%%%%%%%%%%%%%%%%%%%%%%%%

However, it is obvious that the radio source is not identical with
T~Tau~Sb.
Yet they appear to be close enough to imply a relation between
the star and the radio source.
\citet{loinard07a} suggested the radio emission comes from the jets of
Sb hitting the circumbinary disk.
If this is true, then it is surprising that the position of the radio
source relative to T~Tau~Sb changes within a few years.
However, our knowledge of the geometry of the system is limited.
\citet{ratzka09} argue that the circumstellar disk around T~Tau~Sb is
seen not far from face-on, which implies that the jet axis is close
to the line of sight.  The circumbinary disk has to be nearly edge-on
to provide the observed extinction.  This means that the circumbinary
disk is not aligned with the disk around T~Tau~Sb or the binary orbit.
Given this complex geometry, it is conceivable that the offset from
T~Tau~Sb of the intersection of jet and circumbinary disk changes on
the timescale of the binary orbit.

  There appears to be a systematic offset between T~Tau~Sb and the
  radio source, with the radio source always east of the star.
  This might indicate the orientation of the jet.
  However, considering the difficulties in registering the radio
  observations with the infrared images and the uncertainties of the
  T~Tau~N-S orbit which was extrapolated more than 20\,years into the
  past, estimating the jet direction from this systematic offset would
  be extremely uncertain.

Finally, we note that the circumbinary material in front of T~Tau~S
has to be distributed over at least 20 -- 25\,AU in both R.A.\ and
Dec.\ to explain the permanently high extinction of T~Tau~Sb.
The star has been observed from the most southern to the most northern
point in its orbit, but no significant brightening was observed.
The material causing the extinction might not be circumbinary, but a
cloud in the T~Tau system that happens to lie on our line of sight to
T~Tau~S.

%%%%%%%%%%%%%%%%%%%%%%%%%%%%%%%%%%%%%%%%%%%%

\subsection{Are there more components in the system?}

\citet{nisenson85} detected another source $0\farcs27$ north of
T~Tau~N using speckle imaging techniques in the visual wavelength
regime.  The source was later identified in the K-band by
\cite{maihara91}, but neither \cite{gorham92} nor \cite{stapelfeldt98}
detected it.
The magnitude difference to T~Tau~N was $4.33\pm0.09\rm\,mag$ at
521\,nm \citep{nisenson85} and $3.46\pm0.32\rm\,mag$ in the K-band
\citep{maihara91}.

\citet{csepany15} found a tentative companion 144\,mas south of
T~Tau~N.  They found magnitude differences of $4.96\pm0.30\rm\,mag$ at
$1.02\rm\,\mu m$ and $4.37\pm0.30\rm\,mag$ at $1.34\rm\,\mu m$.
Considering the long time difference and the different wavelengths,
this is comparable to the brightness reported by \citet{nisenson85}
and \citet{maihara91}.

If both \citet{nisenson85} and \citet{csepany15} observed
the same object, then its position angle has changed by about
180\degr\ in 30\,years, which corresponds to an orbital period of
60\,years.
With a semi-major axis of $\sim0.2''$ or 30\,AU, a system mass of
$7.5\,M_\odot$ would be required, clearly more than the mass expected
for T~Tau~N plus a faint companion.
It is also not possible that the two sources are a background star.
The proper motion of T~Tau is $(15.51,-13.67)\rm\,mas/yr$
\citep{vanLeeuwen07}, i.e.\ it is moving south.  Therefore, a
background star would move from south to north relative to T~Tau.
We conclude that the two additional companions -- if both are real --
are not the same object.

Unfortunately, the dynamical mass derived by our orbit fit in
Sect.~\ref{OrbitNSSect} cannot be used to exclude additional objects
in the system, since we used our best estimate for the total mass as
constraint for the fit.  It is therefore not surprising that the
total mass of the best orbit is close to the expected total mass.

%%%%%%%%%%%%%%%%%%%%%%%%%%%%%%%%%%%%%%%%%%%%%%%%%%%%%%%%%%%%%%%%%%%%%%%%%%%%%

\section{Summary}
\label{SumSect}

We present an astrometric measurement of the stars in the T~Tauri
triple system obtained with SPHERE, and recalibrated data obtained
with NACO.
After careful calibration, the SPHERE data is as precise as the best
astrometric data from NACO.  Time will tell whether SPHERE is stable
enough to maintain this high precision over several years.

We determine the orbital elements of the T~Tau~Sa/Sb binary and find
very good agreement with the latest orbit by \citet{schaefer14}.
This indicates that the observed fraction of the orbit is large enough
for a good orbit fit.

The orbit of the binary around T~Tau~N is only weakly constrained.
Its period is of the order of several thousand years, therefore no
progress can be expected in the foreseeable future.
However, we successfully use T~Tau~N as astrometric reference to find
the position of the center of mass of the Sa/Sb binary.
This allows us to derive the mass ratio and individual masses of the
stars.
The masses of T~Tau~Sa and Sb are $2.12\pm0.10\,M_\odot$ and
$0.53\pm0.06\,M_\odot$, which confirms that T~Tau~Sa is at least as
massive as T~Tau~N, despite the large contrast in visible light.

We used the newly derived orbits to compute the position of the radio
source reported by \citet{loinard03} in the reference frame of
T~Tau~S.  We find no evidence for a dramatic chance in its orbital
path.  Given its proximity to T~Tau~Sb, the interpretation that it is
related, but not identical to T~Tau~Sb is still valid.

The orientation of the disks in the T~Tau~S binary remains puzzling.
The circumstellar disk around T~Tau~Sb is only moderately inclined
\citep{ratzka09}.
It might be co-planar with the binary orbit, which has an inclination
of about $20^\circ$.
On the other hand, the circumstellar disk around T~Tau~Sa and the
circumbinary disk are seen nearly edge-on, since they have been
suggested to explain the large extinction.
Therefore, both disks are approximately perpendicular to the binary
orbit.
The extinction in front of T~Tau~S is more or less constant over the
observed section of the binary orbit (almost half of it).
Therefore, the material causing the extinction must be distributed
over at least 20 -- 25\,AU.
This raises the question whether it really forms a circumbinary disk,
or is distributed in a different way in front of T~Tau~S.

The orbit of T~Tau~S is now well understood.
Our knowledge of the circumstellar and circumbinary material, however,
remains sketchy.
More high-resolution observations at longer wavelengths will be
required to reveal the orientation of the disks around T~Tau~Sa and
Sb, and the nature of the material in front of both stars.

%%%%%%%%%%%%%%%%%%%%%%%%%%%%%%%%%%%%%%%%%%%%%%%%%%%%%%%%%%%%%%%%%%%%%%%%%%%%%

\acknowledgements

We are grateful to Gail Schaefer for providing her latest astrometric
measurements before publication, which helped to find an error in our
calibration.
We thank the anonymous referee for many suggestions that helped to
improve the paper.
As usual, Friedrich vom Stein was a big help for this work.

\bibliographystyle{bibtex/aa}
\bibliography{TTau2015}

\end{document}